\documentclass[lettersize,journal]{IEEEtran}
\usepackage{amsmath}

\usepackage{epsfig}
\usepackage{amssymb}
\usepackage{tabularx}
\usepackage[skip=1ex, justification=justified]{caption}
\usepackage{multirow}
\usepackage{rotating}
\usepackage{graphicx}
\usepackage{tabularx}
\usepackage{epstopdf}
\usepackage{array}
\usepackage{mathrsfs}
\usepackage{amsthm}
\usepackage{amsfonts}
\usepackage{multicol}
\usepackage{epsfig}
\usepackage{algpseudocode}
\usepackage{csquotes}
\usepackage{float}
\usepackage{url}
\usepackage{blindtext}
\usepackage{cite}
\usepackage{lipsum}
\usepackage{mathtools}
\usepackage{etoolbox}
\usepackage{subcaption}
\usepackage{enumitem}
\usepackage{hyperref}
\usepackage{color}
\usepackage{bm}
\usepackage{float}   
\usepackage[ruled,vlined,linesnumbered,resetcount]{algorithm2e}
\SetKwRepeat{Do}{do}{while}%
\def\BibTeX{{\rm B\kern-.05em{\sc i\kern-.025em b}\kern-.08em
		T\kern-.1667em\lower.7ex\hbox{E}\kern-.125emX}}
\usepackage[T2A,T1]{fontenc}

\makeatletter

\begin{document}
	\title{	Resource Allocation and Beamforming in FIM-Assisted BS and STAR-BD-RIS-Aided NOMA: An AIW-Meta-Learning Approach}
	\author{ Armin Farhadi, Maryam Cheraghy, and Eduard Jorswieck, \IEEEmembership{Fellow,~IEEE}
		\thanks{A. Farhadi is with the School of Electrical and Computer Engineering, College of Engineering, University of Tehran, Tehran, Iran (e-mail: armin.farhadi@ut.ac.ir). 
		M. Cheraghy is with the Department of Computer Science, Wenzhou-Kean University, Wenzhou 325060, China, and also with the Department of Computer Science and Technology, Kean University, Union, NJ 07083 USA (e-mail: mcheragh@kean.edu).  Eduard Jorswieck is with the Institute for Communications Technology,
Technische Universität Braunschweig, 38106 Braunschweig, Germany (email:
jorswieck@ifn.ing.tu-bs.de).
			}}

	\maketitle
{\color{black}
	\begin{abstract}
		
		This paper investigates a flexible intelligent metasurface (FIM)-enabled wireless communication system that integrates simultaneously transmitting and reflecting beyond diagonal reconfigurable intelligent surfaces (STAR-BD-RIS) with non-orthogonal multiple access (NOMA). The considered system consists of a multi-antenna FIM-assisted base station (BS) supported by dual-sector BD-RIS. The FIM is composed of low-cost radiating elements capable of independent signal transmission and dynamic vertical reconfiguration (morphing). The objective is to maximize energy efficiency (EE) by jointly optimizing the BS beamforming, STAR-BD-RIS configuration, NOMA-related variables, and the FIM surface shape under practical power constraints. Due to the highly non-convex nature of the problem, an adaptive inverse-weighted Meta-Soft Actor-Critic (AIW-Meta-SAC) algorithm is proposed. Unlike conventional Meta-SAC approaches, the proposed method employs an adaptive weighting mechanism to effectively incorporate system constraints into the reward function, thereby improving learning efficiency and convergence behavior. Simulation results demonstrate that the proposed AIW-Meta-SAC significantly outperforms the Meta-DDPG baseline. Furthermore, the FIM-assisted STAR-BD-RIS architecture achieves notable energy efficiency gains compared to conventional benchmark schemes.
	\end{abstract}}
	
	
	\section{Introduction} \label{Introduction}

{\color{black}

Next-generation wireless networks are expected to support advanced applications that demand ultra-fast, reliable, and low-latency connections \cite{farhadi2025joint, hengyu2026uav}. Current systems struggle to meet these demands, and traditional solutions raise energy and cost concerns. Metasurfaces have recently emerged as a promising technology to improve both performance and energy efficiency \cite{farhadi2021resource, zavleh2022downlink, javadi2025meta}.


Advances in micro- and nano-fabrication, combined with flexible metamaterials, have led to the creation of flexible intelligent metasurfaces (FIMs), composed of dielectric elements on flexible substrates \cite{an2024downlink}. Unlike rigid metasurfaces, FIMs excel at manipulating waves on curved or irregular surfaces \cite{kamali2016decoupling}. Also, \cite{an2024downlink} investigated the feasibility of using FIM in a communication system.


To enable real-time, programmable 3D shape reconfiguration, \cite{ni2022soft} proposed a soft microfluidic network with liquid metal embedded in elastic material, reshaped via electromagnetic actuation. More recently, \cite{bai2022dynamically} introduced an FIM made of metallic wire arrays controlled by reprogrammable Lorentz forces from electric currents and magnetic fields, enabling fast and precise dynamic morphing.
}

{\color{black}Reconfigurable intelligent surfaces (RISs) play an important role in enhancing both spectral and energy efficiency in future 6G wireless systems. By adjusting the amplitude and phase of incident signals, RISs contribute to improved communication performance and lower energy usage \cite{farhadi2024meta,wu2025intelligent, farhadi2025energy}. However, conventional RISs, which only reflect signals, may experience performance limitations when users fall outside their effective reflection area. To overcome this issue, Simultaneously Transmitting and Reflecting RISs (STAR-RISs) have been proposed. These structures support both transmission and reflection of signals, leading to improved system performance \cite{javadi2025meta}. 

Traditional RISs are typically represented using diagonal phase shift matrices, where each element corresponds to a load-connected unit. In contrast, Beyond Diagonal RIS (BD-RIS) architectures have been introduced, which do not rely solely on diagonal configurations. BD-RIS designs can be divided into three main categories: (1) group- or fully-connected architectures modeled by block-diagonal matrices \cite{farhadi2025meta}, (2) dynamically group-connected structures that adapt based on channel state information (CSI) \cite{farhadi2025meta}, and (3) non-diagonal designs where signal coupling between elements is allowed \cite{farhadi2025meta}. These architectures offer better performance and wider signal coverage compared to standard RIS systems \cite{farhadi2025meta}.
Studies on STAR/BD-RIS have focused on optimization and design, including power-beamforming tradeoffs, energy-efficient transmission, and reconfigurable impedance networks \cite{star-irs-1,NUM8,shen2021modeling}.

Non-orthogonal multiple access (NOMA) has been identified as an effective method for improving spectral efficiency and supporting a large number of users in next-generation wireless networks. Unlike orthogonal schemes such as frequency division multiple access (FDMA), time division multiple access (TDMA), code division multiple access (CDMA), and orthogonal frequency division multiple access (OFDMA), NOMA allows multiple users to share resources by introducing manageable interference, thus enhancing overall network performance \cite{farhadi2021resource, zavleh2022downlink}.

Most existing research applies optimization-based methods for resource allocation in RIS and BD-RIS-assisted networks. Nonetheless, due to the non-convex nature of these problems, such methods often yield suboptimal solutions. Deep Reinforcement Learning (DRL) has emerged as an alternative, although its effectiveness is generally limited to static scenarios \cite{de2025gwo, zarini2025unmanned}. To better address dynamic environments typical of future wireless networks, recent approaches integrate meta-learning with DRL \cite{farhadi2025joint,farhadi2025meta}. This enhances adaptability and performance in dynamic environments, making it well-suited for beyond-5G and 6G networks.\cite{Meta}.}

Most of the previously published papers on FIM investigated the feasibility of using FIM in the new generation of wireless systems, while only a few of these studies addressed resource allocation optimization through conventional methods and different system models \cite{eftekhari2026hybrid,an2024downlink,kamali2016decoupling}. In this paper, we investigate a system model that employs a FIM-equipped base station (BS) and a STAR-BD-RIS to enhance user communication using NOMA as the multiple access scheme. Accordingly, we formulate an optimization problem to maximize energy efficiency (EE), considering NOMA, FIM, the maximum available power at the BS, and the STAR-BD-RIS constraints. To address this issue, we adopt the soft actor-critic (SAC) algorithm due to its efficiency in handling continuous action spaces with relatively low computational complexity. To further incorporate the benefits of meta-learning, we extend SAC by integrating it with a meta-learning framework, resulting in the proposed Meta-SAC algorithm. Proper incorporation of constraints into the reward function in reinforcement learning is critically important and plays a pivotal role in ensuring valid and reliable responses. Existing works \cite{farhadi2024meta, javadi2025meta, farhadi2025joint} consider fixed weighting factors to account for constraints in the reward function; however, this approach requires extensive tuning of these weights. To address this limitation, we enhance the learning-based solution by introducing an adaptive inverse weighting scheme within the Meta-SAC framework, referred to as AIW-Meta-SAC, which enables efficient and dynamic incorporation of constraints into the reward function. Moreover, to manage optimization problems involving both continuous and discrete decision variables, we apply a quantization approach, which enables efficient problem-solving while maintaining a balance between solution quality and computational cost. Also, simulation results demonstrate that the FIM-based BS achieves higher energy efficiency compared to systems assisted by conventional BS. Furthermore, the proposed Meta-SAC algorithm shows superior performance across various system parameters, outperforming the meta deep deterministic policy gradient (Meta‑DDPG) method \cite{Meta}.

Our main contributions in this paper are summarized as below:
\begin{itemize}
    \item We propose a novel FIM-assisted communication system integrated with a STAR-BD-RIS and NOMA scheme to enhance the performance of next-generation wireless networks.

    \item We formulate an EE maximization problem by jointly considering beamforming design, resource allocation, transmit power constraints at the BS, and STAR-BD-RIS constraints.

    \item We develop a reinforcement learning-based solution by adopting the SAC algorithm and further extend it with a meta-learning framework, resulting in the proposed Meta-SAC algorithm.

    \item We introduce an adaptive inverse weighting mechanism, referred to as AIW-Meta-SAC, which dynamically adjusts constraint penalties in the reward function and eliminates the need for manual weight tuning.

    \item To efficiently handle mixed discrete and continuous optimization variables, we employ a quantization-based approach that achieves a favorable trade-off between computational complexity and performance.

    \item Simulation results demonstrate that the proposed FIM-based system significantly improves EE compared to conventional BS architectures, while the proposed Meta-SAC framework outperforms benchmark methods such as Meta-DDPG.
\end{itemize}

Throughout this paper, we follow standard notation conventions. Matrices are represented by bold uppercase letters, vectors by bold lowercase letters, and scalars by regular lowercase letters. The identity matrix is denoted by $\mathbf{I}$. A complex-valued matrix of size $x \times y$ is expressed as $\mathbb{C}^{x \times y}$, and the superscript $H$ denotes the conjugate transpose of a matrix. Moreover, $\mathbb{E}[\cdot]$ indicates the expectation operator, while $\mathrm{diag}(\cdot)$ represents a diagonal matrix. The gradient of a function $f(\cdot)$ is denoted by $\nabla f(\cdot)$, and the Euclidean (2-) norm of a vector $\mathbf{s}$ is written as $\|\mathbf{s}\|$.

The remainder of this paper is organized as follows. Section \ref{System, Channel, and Signal Models} presents the system, channel and signals model under consideration. In Section \ref{Optimization Problem}, the optimization problem is formulated, focusing on transmit beamforming as well as phase-shift and amplitude coefficient design, while considering NOMA-assisted FIM in a multiple‑input and single‑output (MISO) system. Section \ref{Proposed Meta-SAC Algorithm} introduces the proposed AIW-Meta-reinforcement learning approach, which is applied to jointly optimize transmit beamforming, phase shift matrix of the STAR-BD-RIS, and NOMA related optimization variables in a multi-user MISO scenario. The performance of the proposed method is evaluated through simulation results in Section \ref{Simulation result}. Finally, Section \ref{Conclusion} concludes the paper and summarizes the main findings and contributions.

{\color{black}\section{System, Channel, and Signal Models}\label{System, Channel, and Signal Models}
\subsection{System Model}

As illustrated in Fig. \ref{sysmodel}, we examine the downlink of a multiuser multiple-input single-output (MISO) system in a single-cell, where a BS equipped with a FIM serves $U$ single-antenna users concurrently. The FIM antenna structure is assumed to be a uniform planar array (UPA) positioned on the $x$-$z$ plane. Let $M = M_x M_z \geq U$ be the total number of transmit antennas, where $M_x$ and $M_z$ refer to the number of antenna elements along the $x$-axis and $z$-axis, respectively. The sets of antennas and users are defined as $\mathcal{M} \triangleq \{1, 2, \ldots, M\}$ and $\mathcal{U} = \{1, \dots, U_t, U_t+1, \dots, U_t+U_r\} $. The STAR-BD-RIS divides its coverage area into two parts: a transmission region, which serves \( U_t \) users, and a reflection region, which serves \( U_r \) users. Let $\alpha_{u,n} \in \{0,1\}$ be a binary variable that represents whether subcarrier $n$ is assigned to user $u$. Specifically, $\alpha_{u,n} = 1$ means subcarrier $n$ is assigned to user $u$, and $\alpha_{u,n} = 0$ means it is not assigned.

\begin{figure}[t!]
	\centering
	\includegraphics[width=1
	\linewidth]
	{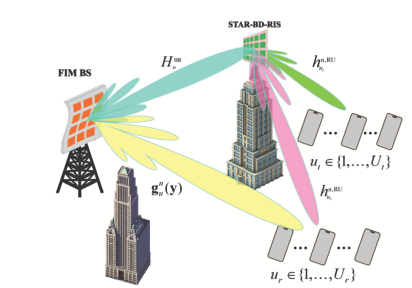}
	\caption{ Communication system model with BS equipped with FIM and STAR-BD-RIS for multi-user.}%
	\label{sysmodel}%
	\vspace{8pt}
\end{figure}


Also, we consider a communication system assisted by a STAR-BD-RIS, which supports two sectors, similar to the STAR-RIS structure, as illustrated in Fig.~\ref{sysmodel}. In scenarios where the direct link between the BS and users is obstructed by physical barriers such as buildings, a STAR-BD-RIS is employed to facilitate communication. The STAR-BD-RIS is equipped with \( K_{\text{RIS}} \) antenna elements, indexed by the set \( \mathcal{K}_{\text{RIS}} = \{1, 2, \dots, K_{\text{RIS}}\} \).

In the investigated system model, the FIM-equipped BS divided the total available bandwidth into a set of subcarriers indexed by \( \mathcal{N} = \{1, 2, \ldots, N\} \). To support enhanced communication, the system adopts a NOMA scheme.

Unlike traditional systems with fixed antenna arrays, each radiating unit of the FIM can be dynamically adjusted in the vertical ($y$-axis) direction via a controller. Let $\mathbf{q}_m = [x_m, y_m, z_m]^T \in \mathbb{R}^3$ denote the position of the $m$-th radiating unit, where $m \in \mathcal{M}$. The coordinates along the $x-$ and $z-$axes are defined as $x_m = d_x \times \bmod(m-1, M_x)$ and $z_m = d_z \times \left\lfloor \frac{m-1}{M_x} \right\rfloor,$ respectively.

for all $m \in \mathcal{M}$, where $d_x$ and $d_z$ represent the inter-element spacing along the $x$- and $z$-axes, respectively. The vertical coordinate $y_m$ of each element is variable within a predefined range due to the flexibility of the metasurface, satisfying $ y_{\text{min}} \leq y_m \leq y_{\text{max}}, \quad \forall m \in \mathcal{M} $
, here, $y_{\text{min}}$ and $y_{\text{max}}$ indicate the minimum and maximum heights allowed by the FIM's deformable design, and the morphing range is defined as $\Delta y = y_{\text{max}} - y_{\text{min}} > 0$. In this work, we assume $y_{\text{min}} = 0$ without loss of generality. The deformable profile of the FIM can be expressed as $ \mathbf{y} = [y_1, y_2, \ldots, y_M]^T \in \mathbb{R}^{M\times 1}. $



We assume a quasi-static flat fading environment for all wireless channels. Let $\mathbf{g}_u^n \in \mathbb{C}^{M \times 1}$ denote the complex baseband channel from the BS's FIM to the $u$-th user, for $u \in \mathcal{U}$ and $n$th subcarrier. The channel follows a multipath model incorporating contributions from multiple propagation paths.

\sloppy For each scattering component observed in the far field, the phase response across the antenna array depends on the elevation angle $\theta \in [0, \pi)$, azimuth angle $\varphi \in [0, \pi)$, and the surface configuration vector $\mathbf{y}$. The array steering vector $ \mathbf{v}(\mathbf{y}, \varphi, \theta)$ is then expressed as 
\begin{align*}
\mathbf{v}(\mathbf{y}, \varphi, \theta) = \big[ &1, \ldots, e^{j \omega (x_m \sin\theta \cos\varphi + y_m \sin\theta \sin\varphi + z_m \cos\theta)}, \\
&\ldots, e^{j \omega (x_M \sin\theta \cos\varphi + y_M \sin\theta \sin\varphi + z_M \cos\theta)} \big]^T
\end{align*}
where \( \omega = \frac{2\pi}{\kappa} \) represents the wave number, and \( \kappa \) denotes the carrier wavelength.

Moreover, consider that each user communicates with the BS through $P$ distinct propagation paths, indexed by the set $\mathcal{P} \triangleq \{1, 2, \ldots, P\}$. The complex channel gain associated with the $p$-th path for user $u$ is denoted by $\gamma_{u, p}^n \in \mathbb{C}, \ \forall p \in \mathcal{P}, \forall u \in \mathcal{U}, \forall n \in \mathcal{N}$. The angles of arrival for the $p$-th path are defined by its elevation $\vartheta_p$ and azimuth $\varphi_p$. Consequently, the narrowband wireless channel from the BS to user $u$ on $n$th subcarrier can be expressed as $\mathbf{g}_u^n(\mathbf{y}) = \sum_{p=1}^{P} \gamma_{u, p}^n \mathbf{v}(\mathbf{y}, \varphi_p, \vartheta_p), \quad \forall u \in \mathcal{U}, \quad \forall n \in \mathcal{N}.$

In this work, we assume that the coefficients $\gamma_{u, p}^n \in \mathbb{C}, \ \forall p \in \mathcal{P}, \forall u \in \mathcal{U}, \forall n \in \mathcal{N}$ are independent and identically distributed (i.i.d.) circularly symmetric complex Gaussian (CSCG) random variables, following the distribution $\gamma_{u, p}^n \sim \mathcal{CN}(0, (\sigma_{u, p}^n)^2)$, where $(\sigma_{u, p}^n)^2$ is the average power of the $p$-th path for user $u$ and subchannel~\cite{heath2016overview}. The overall large-scale fading (path loss) for user $u$ is denoted by $\eta_u^n$, which satisfies the $\sum_{p=1}^{P} (\sigma_{u, p}^n)^2 = \eta_u^n, \quad \forall u \in \mathcal{U}, \quad \forall n \in \mathcal{N}.$

The same formulations and explanations that apply to the channel from FIM-equipped BS to users also apply here. Let \( \mathbf{H}_n^{BR}(\mathbf{y}) \in \mathbb{C}^{ M \times K_{\text{RIS}}} \) denote the narrowband channel matrix between the FIM-equipped BS and the STAR-BD-RIS in $n$th subcarrier. The complex path gain for the \( p \)-th path is represented by \( \gamma_{p}^{k, n} \in \mathbb{C} \). The departure angles at the RIS side are denoted by \( \phi^{(\text{k})}_p \) and \( \psi^{(\text{k})}_p \).

Under the far-field assumption, the FIM BS-to-RIS channel vector for $k$th element can be modeled as $\mathbf{h}^{BR}_{k, n} (\mathbf{y})  = \sum_{p=1}^{L} \gamma_{p}^{k, n} \, \mathbf{a}_k(\mathbf{y}, \phi^{(\text{k})}_p, \psi^{(\text{k})}_p), \quad \forall k \in \mathcal{\mathcal{K}_{\text{RIS}}}$
, where \( \mathbf{a}_k(\mathbf{y}, \phi^{(\text{k})}_p, \psi^{(\text{k})}_p) \in \mathbb{C}^{M \times 1} \) denotes the FIM's deformable steering vector, as defined earlier for the $k$th STAR-BD-RIS and $L$ is number of propagation paths.

A dual-sector BD-RIS can be represented using two matrices \( \boldsymbol{\Phi}^s \in \mathbb{C}^{K_{\text{RIS}} \times K_{\text{RIS}}} \), where \( s \in \mathcal{S} \) denotes the sector index. Each matrix captures the signal transformation behavior of one sector. These matrices are extracted as sub-matrices from a higher-dimensional scattering matrix \( \boldsymbol{\Phi} \in \mathbb{C}^{2K_{\text{RIS}} \times 2K_{\text{RIS}}} \), which models a reconfigurable impedance network with \( 2K_{\text{RIS}} \) ports. Specifically, each matrix \( \boldsymbol{\Phi}^s \) is defined as $ \boldsymbol{\Phi}^s = [\boldsymbol{\Phi}]_{(K_{\text{RIS}}+1 : 2 K_{\text{RIS}},\; 1 : K_{\text{RIS}})} $

The matrices \( \boldsymbol{\Phi}^s \) satisfy a joint unitary constraint given by $ \sum_{s \in \mathcal{S}} (\boldsymbol{\Phi}^s)^H \boldsymbol{\Phi}^s = \mathbf{I}_{K_{\text{RIS}}}. $
For additional insights into the modeling, design, and performance analysis of multi-sector BD-RIS systems, refer to \cite{farhadi2025meta}.

As discussed in \cite{farhadi2025meta}, reconfigurable impedance networks with different circuit topologies result in varying mathematical structures for their scattering matrices. In this work, we consider a \textit{cell-wise single-connected (CW-SC)} architecture for the \textit{multi-sector BD-RIS}. For better illustration, Fig.~\ref{BD_RIS_structure} shows the proposed configuration. In this architecture, the antennas within each cell are interconnected through reconfigurable impedance components, while antennas across different cells remain unconnected.

\begin{figure}[t]
	\centering
	\includegraphics[width=0.8
	\linewidth]
	{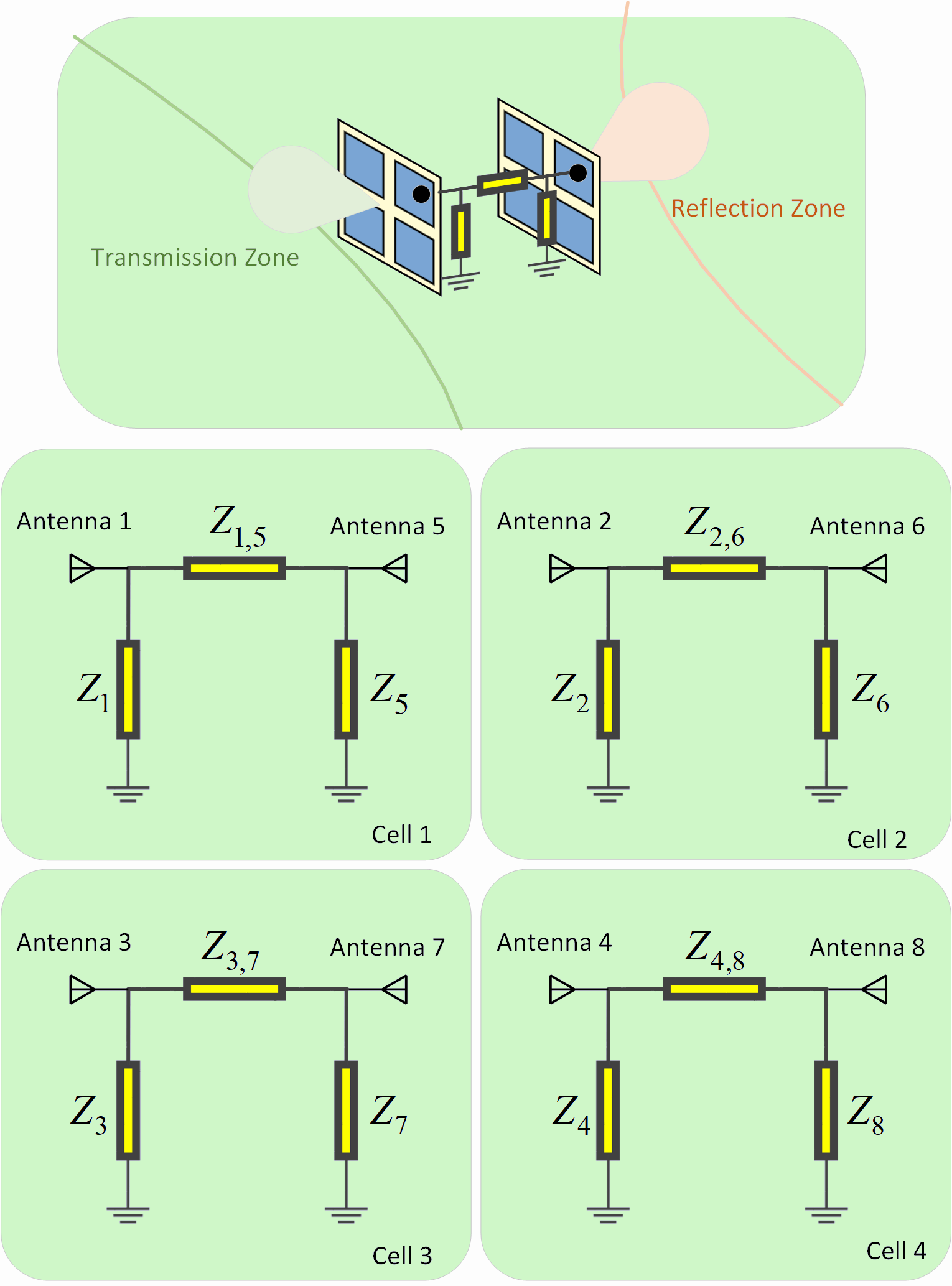}
	\caption{The design of a cell-wise single-connected BD-RIS consists of two sectors and four individual cells, with each cell containing a pair of antennas.}%
	\label{BD_RIS_structure}%
	\vspace{8 pt}
\end{figure} 

Accordingly, each sector-specific matrix \( \boldsymbol{\Phi}^s \in \mathbb{C}^{K_{\text{RIS}} \times K_{\text{RIS}}} \), extracted as a sub-matrix from the global scattering matrix \( \boldsymbol{\Phi} \in \mathbb{C}^{2K_{\text{RIS}} \times 2K_{\text{RIS}}} \), becomes \textit{diagonal} under the CW-SC architecture. For each sector \( s \in \mathcal{S} \), the matrix \( \boldsymbol{\Phi}^s \) is given by $ \boldsymbol{\Phi}^s = \operatorname{diag} \left( \Phi^{s,1}, \ldots, \Phi^{s,K_{\text{RIS}}} \right), $
where \( \Phi^{s,k} \in \mathbb{C} \) for all \( k \in \{1, \ldots, K_{\text{RIS}}\} \) and \( s \in \mathcal{S} \). Consequently, the joint unitary constraint $ \sum_{s \in \mathcal{S}} (\boldsymbol{\Phi}^s)^H \boldsymbol{\Phi}^s = \mathbf{I}_{K_{\text{RIS}}} $
reduces to a per-element power constraint of the form 
\[
\sum_{s \in \mathcal{S}} \left| \Phi^{s,k} \right|^2 = 1, \quad \forall k \in \{1, \ldots, K_{\text{RIS}}\}.
\]

Let \( \mathbf{h}_{u}^{\text{n, RU}} = \left[ h_{u,1}^{\text{n, RU}}, \ldots, h_{u,K_{\text{RIS}}}^{\text{n, RU}} \right]^T \in \mathbb{C}^{K_{\text{RIS}} \times 1} \) denote the channel vector from the STAR-BD-RIS to the \( u \)-th user and $n$th subcarrier, where \( u \in \mathcal{U}_s \) and \( s \in \mathcal{S} = \{\text{T}, \text{R}\} \) represents the sector index, with \( \text{T} \) and \( \text{R} \) corresponding to the transmission and reflection regions, respectively.

The signal-to-interference-and-noise ratio (SINR) for user $u$ on subcarrier $n \in \mathcal{N}$ is expressed as:
\begin{align}
&\Gamma_u^n =\\ \nonumber &\frac{\alpha_{u,n} 
    \left| \left(  (\bm{h}_{u}^{\mathrm{n,RU}})^ \mathrm{H} \boldsymbol{\Phi}^s (\bm{H}_n^{\mathrm{BR}})^ \mathrm{H} + (\bm{g}_u^{n})^\mathrm{H} \right)\bm{w}_u^n  \right|^2
}{
    \sum_{\substack{i \in \mathcal{U} \\ i \neq u}}\alpha_{i,n}  \left| \left(  (\bm{h}_{u}^{\mathrm{n,RU}})^ \mathrm{H} \boldsymbol{\Phi}^s (\bm{H}_n^{\mathrm{BR}})^ \mathrm{H} + (\bm{g}_u^{n})^\mathrm{H}  \right)\bm{w}_i^n \right|^2 + (\sigma_u^n)^2
},
\end{align}
where $\bm{w}_u^n$ denotes the beamforming vector for the $u$th user and $n$th subcarrier. The total achievable rate is given by 
\begin{equation}
R_T = \sum_{u \in \mathcal{U}} \sum_{n \in \mathcal{N}} \log_2(1 + \Gamma_u^n).
\end{equation}
The total power consumption consists of $ P_T = P_{T_1} + P_{T_2}, $
where $ P_{T_1} = P_{\text{St}}^{\text{BS}} + P_{\text{BD-RIS}}, \quad P_{\text{BD-RIS}} = P_{\text{St}}^{\text{BD-RIS}} + K_{\text{RIS}} P_{\text{Dn}}^{\text{BD-RIS}}, $
and $P_{T_2} = \frac{1}{\eta}\sum_{u \in \mathcal{U}} \sum_{n \in \mathcal{N}}\alpha_{u,n}   \|\bm{w}_u^{n} \|^2$, here, $\eta \in (0,1)$ is the power amplifier efficiency.
\section{Optimization Problem}\label{Optimization Problem}
The goal is to maximize the system's EE, defined as $ \mathrm{EE} = \frac{R_T}{P_T}. $
The corresponding optimization problem is formulated as:
\begin{subequations}\label{OptProb}
	\begin{align}
	&\text{OP:}~\max_{ \{\mathbf{y} , \bm{w}_u^n,\, \alpha_{u,n},\, \boldsymbol{\Phi}^s\}} \quad \mathrm{EE} \label{Obj} \\[2pt]
	\text{s.t.} \quad 
	& \Gamma_{u \rightarrow i}^n - \Gamma_u^n \geq 0, 
	\quad \forall n \in \mathcal{N} ,\, u, i \in  \mathcal{U},\, u \neq i, \label{C1} \\
	& \sum_{u \in \mathcal{U}} \alpha_{u,n} \leq U_{\max}, 
	\quad \forall n \in \mathcal{N}, \label{C2} \\
	& \sum_{u \in \mathcal{U}} \alpha_{u,n} \|\bm{w}_u^n\|^2 \leq P_{\max}, 
	\quad \forall n \in \mathcal{N}, \label{C3} \\
	&y_{\text{min}} \leq y_m \leq y_{\text{max}}, \quad \forall m \in \mathcal{M} , \label{C4} \\
	& \alpha_{u,n} \in \{0,1\}, 
	\quad \forall u \in \mathcal{U},\, \forall n \in \mathcal{N}, \label{C5} \\
	& \sum_{s \in \mathcal{S}} \left| \Phi^{s,k} \right|^2 = 1, 
	\quad \forall k \in \mathcal{K}_{\text{RIS}}. \label{C6}
	\end{align}
\end{subequations}
Constraint \eqref{C1} guarantees that each user can successfully perform successive interference cancellation (SIC). Constraint \eqref{C2} limits the number of users that can share the same subcarrier to a maximum of $U_{\max}$. Additionally, constraint \eqref{C3} represents the maximum available power at the FIM-BS. Also, \eqref{C4} shows the deformation of each FIM surface element to remain within
the physical limits. Constraint \eqref{C5} imposes the binary condition on the subcarrier assignment variable. Finally, constraint \eqref{C6} defines the STAR-BD-RIS configuration condition.

The presence of binary variables and non-convex expressions makes this a mixed-integer nonlinear programming (MINLP) problem. To tackle this challenge, we propose a \textit{Meta-SAC algorithm} that combines soft actor-critic reinforcement learning with meta-learning techniques for efficient and robust optimization.

}

{\color{black}\section{Proposed AIW-Meta-SAC Algorithm}\label{Proposed Meta-SAC Algorithm}
This section presents a detailed description of the proposed  AIW-Meta-SAC algorithm. The optimization problem in \eqref{OptProb} is reformulated as a Markov Decision Process (MDP) defined by the state space $\mathbb{S}$, action space $\mathcal{A}$, and reward function $\mathcal{R}$.

\subsubsection{State Space $\mathbb{S}$}
The state space $\mathbb{S}$ includes all relevant system information, such as user channels, {\color{black}interference levels, and the total data rate $R_T$. It is formally expressed as:
\[
\mathbb{S} = \left\{
\{ \bm{h}_{u}^{\mathrm{n,RU}}, \bm{g}_u^{n}, \Gamma_u^n \}_{\forall n \in \mathcal{N}, \, \forall u \in \mathcal{U}}, \;
\{ \bm{H}_n^{\mathrm{BR}} \}_{\forall n \in \mathcal{N}}, \;
R_T
\right\}.
\]

}

\subsubsection{Action Space $\mathcal{A}$}
The action space $\mathcal{A}$ is composed of all decision variables involved in the optimization problem \eqref{OptProb}:
\[
\mathcal{A} = \left\{ \mathbf{y} , \{\bm{w}_u^n,\, \alpha_{u,n}\}_{\forall u \in \mathcal{U}, \forall n \in \mathcal{N}}, \{\boldsymbol{\Phi}^s\}_{\forall s \in \mathcal{S}} \right\}.
\]

\subsubsection{Reward Function $\mathcal{R}$}
The reward function $R$ is designed to incorporate both the optimization objective and penalties for constraint violations. It is defined as:
{\color{black}

\begin{align}
r = {} 
& \nu_{1} \underbrace{EE}_{(f_1)} 
+ \nu_{2} \underbrace{\sum_{u, i \in \mathcal{U}, u \neq i} \sum_{n = 1}^{\mathcal{N}} \left( \Gamma_u^n(i) - \Gamma_u^n(u) \right)}_{(f_2)} \nonumber \\
& + \nu_{3} \underbrace{\sum_{n = 1}^{\mathcal{N}} \left( U_{\max} - \sum_{u \in \mathcal{U}} \alpha_{u,n} \right)}_{(f_3)} \nonumber 
\end{align}
\begin{align}  \label{reward_fn_numbered}
& + \nu_{4} \underbrace{\left( P_{\max} - \sum_{n = 1}^{\mathcal{N}} \sum_{u \in \mathcal{U}} \alpha_{u,n} \| \bm{w}_u^n \|^2 \right)}_{(f_4)} \nonumber \\
& - \nu_{5} \underbrace{\left| \sum_{k \in \mathcal{K}_{\text{RIS}}} \sum_{s \in \mathcal{S}} \left( \left| \Phi^{s,k} \right|^2 - 1 \right) \right|}_{(f_5)}.
\end{align}

The final reward is:
\begin{equation}\label{reward2}
R=
\begin{cases}
r, & \text{if constraint \eqref{C4} --\eqref{C5} is satisfied},\\
-|r|, & \text{otherwise}.
\end{cases}
\end{equation}

The reward function incorporates penalties using weighted factors $\nu_i$ to normalize the terms with different physical units, satisfying $\sum_{i = 1}^{5} \nu_{i} = 1$ and $\nu_{i} \in [0, 1]$ \cite{farhadi2025joint}. } As mentioned in the \ref{Introduction} section, the previous works utilized fixed weighted factors to involve the optimization problem constraints in the proposed meta-learning-based solution. In this case, we should execute the simulation several time to optimize and tune the weighted factors. This procedure is time consuming and it is not suitable for next generation optimization problem. In this paper we proposed a AIW mechanism to adaptively adjust weighted factors. The adaptive weighting coefficients are computed based on the historical value of each term $f_i, \quad \text{for } i \in \{1, \dots, 5\}$
in defined reward in  \eqref{reward_fn_numbered}. Specifically, each weighting factor is defined as
\begin{equation}
\nu_i(t) = \frac{\left( \displaystyle \frac{1}{\bar{f}_i(t)+\epsilon} \right)}
{\displaystyle \sum_{j=1}^{5} \left( \frac{1}{\bar{f}_j(t)+\epsilon} \right)}, \quad i = \{1, \dots, 5\},
\end{equation}
where $\epsilon$ is a small value and $\bar{f}_i(t)$ can be express as 
\begin{equation}
\bar{f}_i(t) = \frac{1}{t} \sum_{\tau=1}^{t} \left| f_i(\tau) \right|,  \quad  i = \{1, \dots, 5\},
\label{running average}
\end{equation}
represents the running average of the absolute value of the $i$-th term up to time step $t$.

\subsection{Meta-SAC Framework}
Meta-SAC extends the SAC approach, incorporating actor and critic networks with parameters $\theta^{\pi}$ and $\theta^{V}$, respectively. Target networks with parameters $\theta^{\pi'}$ and $\theta^{V'}$ are also used to stabilize training.

The objective is to find an optimal policy $\pi^{*}$ that maximizes both the expected cumulative reward and entropy:
\begin{align}
\pi^{*} &= \arg \max_{\pi} \, \mathbb{E}_{(s_n, a_n) \sim \mathbb{P}} \nonumber \\
&\quad \left[ \sum_{n=0}^{\infty} \gamma^n V(s_n, a_n \mid \theta^{V}) - \lambda \sum_{n=0}^{\infty} \gamma^n \log \pi(s_n \mid \theta^{\pi}) \right]
\end{align}

where $\gamma$ denotes the discount factor and $\mathbb{P}$ represents the transition probabilities of the environment.

The critic network is updated by minimizing 
\begin{align} \label{critic-loss}
&L^{\mathrm{critic}}_{\mathrm{SAC}} = \\ \nonumber
& \mathbb{E}_{s \sim p_\pi} \left[ \alpha \log \pi(s \mid \theta^{\pi}) - V(s, a \mid \theta^{V}) \mid a = \pi(s \mid \theta^{\pi}) \right].
\end{align}
Meta-learning is formulated as a bi-level optimization problem, where the actor adapts quickly to new tasks while the meta-critic guides long-term generalization\cite{farhadi2025meta}:
\begin{equation} \label{meta-bilevel}
\begin{aligned}
\eta &= \arg \min_{\eta} \, L^{\mathrm{meta}}(D_{\mathrm{val}}; \theta^{\pi}) \\
\text{subject to} \quad \theta^{\pi^*} &= \arg \min_{\theta^{\pi}} \left( J(D_{\mathrm{trn}}; \theta^{\pi}) \right. \\
&\quad \left. + L^{\mathrm{meta\text{-}critic}}_{\eta}(D_{\mathrm{trn}}; \theta^{\pi}) \right),
\end{aligned}
\end{equation}

where $\eta$ represents the meta-knowledge, updated as follows:
\begin{equation}
\theta^{\pi}_{\mathrm{old}} \leftarrow \theta^{\pi} - \mathrm{lr}_{\mathrm{actor}} \nabla_{\theta^{\pi}} J(\theta^{\pi}), \label{eq:old}
\end{equation}
\begin{equation}
\theta^{\pi}_{\mathrm{new}} \leftarrow \theta^{\pi}_{\mathrm{old}} - \mathrm{lr}_{\mathrm{actor}} \nabla_{\theta^{\pi}} L^{\mathrm{meta\text{-}critic}}_{\eta}. \label{eq:new}
\end{equation}
The actor and critic parameters are updated using:
\begin{equation} \label{actor-parameters}
\theta^{\pi} \leftarrow \theta^{\pi} - \mathrm{lr}_{\mathrm{actor}} \left( \nabla_{\theta^{\pi}} J(\theta^{\pi}) + \nabla_{\theta^{\pi}} L^{\mathrm{meta\text{-}critic}}_{\eta} (\theta^{\pi}) \right),
\end{equation}
\begin{equation} \label{critic-parameters}
\theta^{V} \leftarrow \theta^{V} - \mathrm{lr}_{\mathrm{critic}} \nabla_{\theta^{V}} L(\theta^{V}).
\end{equation}

Meta-knowledge $\eta$ is updated via 
\begin{equation} \label{meta-knowledge-updating}
\eta \leftarrow \eta - \mathrm{lr}_{\mathrm{meta}} \nabla_{\eta} L^{\mathrm{meta}}(\theta^{\pi}).
\end{equation}
\subsection{Meta-SAC Algorithm}

During each time step, the agent observes state $s_t$, selects action $a_t$ based on the policy $\pi$, receives reward $r_t$, and transitions to the next state $s_{t+1}$. Transitions are stored for training. Parameters are updated using training and validation mini-batches as described. The AIW-Meta-SAC algorithm was presented  in \textbf{Algorithm 1}.
\subsection{Computational Complexity of Meta-Learning}

Following \cite{farhadi2024meta}, the computational complexity of Meta-SAC is given by:
\begin{equation}
\begin{split}
\mathcal{O} \left( \sum_{\ell=0}^{L_{\mathrm{actor}} - 1} \nu^{\mathrm{actor}}_{\ell} \nu^{\mathrm{actor}}_{\ell+1} + \sum_{k=0}^{L_{\mathrm{critic}} - 1} \nu^{\mathrm{critic}}_{k} \nu^{\mathrm{critic}}_{k+1} \right. \\
\left. + \sum_{m=0}^{L_{\mathrm{meta}} - 1} \nu^{\mathrm{meta}}_{m} \nu^{\mathrm{meta}}_{m+1} \right),
\end{split}
\end{equation}

where $L_{\mathrm{actor}}$, $L_{\mathrm{critic}}$, and $L_{\mathrm{meta}}$ denote the number of layers in actor, critic, and meta-critic networks, respectively, and $\nu$ represents neurons per layer.

}

\begin{algorithm}[t] \label{algorithm1}
\footnotesize
\SetKwInOut{Input}{Input}
\SetKwInOut{Output}{Output}

\Input{Max episodes $E_{\max}$, time steps $T_{\max}$, gradient steps $G_{\max}$}
\Output{Optimized policy $\pi^{*}$ and meta-knowledge $\eta$}

Initialize replay buffer $\mathcal{D}$ \\
Initialize actor $\pi(s|\theta^{\pi})$, critic $V(s,a|\theta^{V})$ \\
Initialize target networks $\pi'(s|\theta^{\pi'}), V'(s,a|\theta^{V'})$ \\
Initialize meta-knowledge $\eta$ \\
Initialize running averages $\bar{f}_i(0)=f_i$, $i=\{1,\dots,5\}$ \\

\For{each episode $e=1,\dots,E_{\max}$}{
    Observe initial state $s_0$ \\

    \For{each time step $t=1,\dots,T_{\max}$}{
        Select action $a_t \sim \pi(\cdot|s_t)$ \\
        Execute $a_t$, observe $r_t$, $s_{t+1}$ \\
        Decompose reward into components $f_i(t), i=\{1,\dots,5\}$ \\
        Store transition $(s_t,a_t,r_t,s_{t+1})$ in $\mathcal{D}$ \\

        Update running averages with \ref{running average}

        Compute adaptive weights (AIW):
        \[
        \nu_i(t) = \frac{\left(\frac{1}{\bar{f}_i(t)}\right)}
        {\sum_{j=1}^{5}\left(\frac{1}{\bar{f}_j(t)}\right)} , \quad i = \{1, \dots, 5\},
        \]
    }

    \For{each gradient step $n=1,\dots,G_{\max}$}{
        
        Sample mini-batch $D_{\mathrm{trn}}$ from $\mathcal{D}$ \\

        \textbf{Critic update:} $\theta^{V}  \text{using } \eqref{critic-parameters}$

        Compute target policy actions using $\pi'$ and compute temporal-difference targets \\

        \textbf{Actor meta-step construction:}

        Compute $\theta^{\pi}_{\mathrm{old}}$ and $\theta^{\pi}_{\mathrm{new}}$ via \eqref{eq:old} and \eqref{eq:new} \\
        
        Sample validation batch $D_{\mathrm{val}}$ \\

        \textbf{Meta-loss with AIW weighting:}

        Update meta-knowledge $\eta$ via \eqref{eq:old} and \eqref{eq:new}    }
}
\caption{Proposed Meta-SAC Algorithm with AIW Mechanism}
\end{algorithm}

{\color{black}\section{Simulation results}\label{Simulation result}
	In this section, we numerically evaluate the performance of our proposed AIW-Meta-SAC-based beamforming and resource allocation framework in the STAR-BD-RIS-aided NOMA-based wireless network, where the BS is equipped with a FIM. To this end, it is assumed that the BS is located at the coordinate origin, and a STAR-BD-RIS are placed at a distance from it. In addition, all users are located in a circle according to uniform random distribution. All simulation parameters and hyper-parameters of AIW-Meta-SAC are given in Table I unless stated otherwise. 
	
	The performance evaluation is conducted through three benchmarks. The first benchmark investigates the impact of the actor learning rate on system performance. The second benchmark analyzes the effect of varying the number of users. Finally, the proposed model is compared with the conventional diagonal RIS (D-RIS) and alternative meta-learning approaches to assess its effectiveness.

	First, we investigate the learning behavior of the AIW-Meta-SAC algorithm by considering different actor network learning rates. Since the learning rate plays a crucial role in the training of neural networks, its proper tuning is essential for achieving reliable performance in our problem. As illustrated in Fig. \ref{Fig1}, the learning rate specified in Table I has been carefully selected to ensure effective convergence. For comparison, actor learning rates of $0.99$, $10^{-1}$, and $10^{-3}$ are also evaluated, and their learning curves demonstrate that these values are unsuitable, as the agent fails to learn the optimal strategy through interaction with the environment. Moreover, excessively small learning rates significantly increase the number of steps required for convergence, as observed in the initial episodes of the learning curves.

	\begin{figure}[!b]
   \vspace{5pt}
    \centering
    \includegraphics[width=0.4\textwidth]{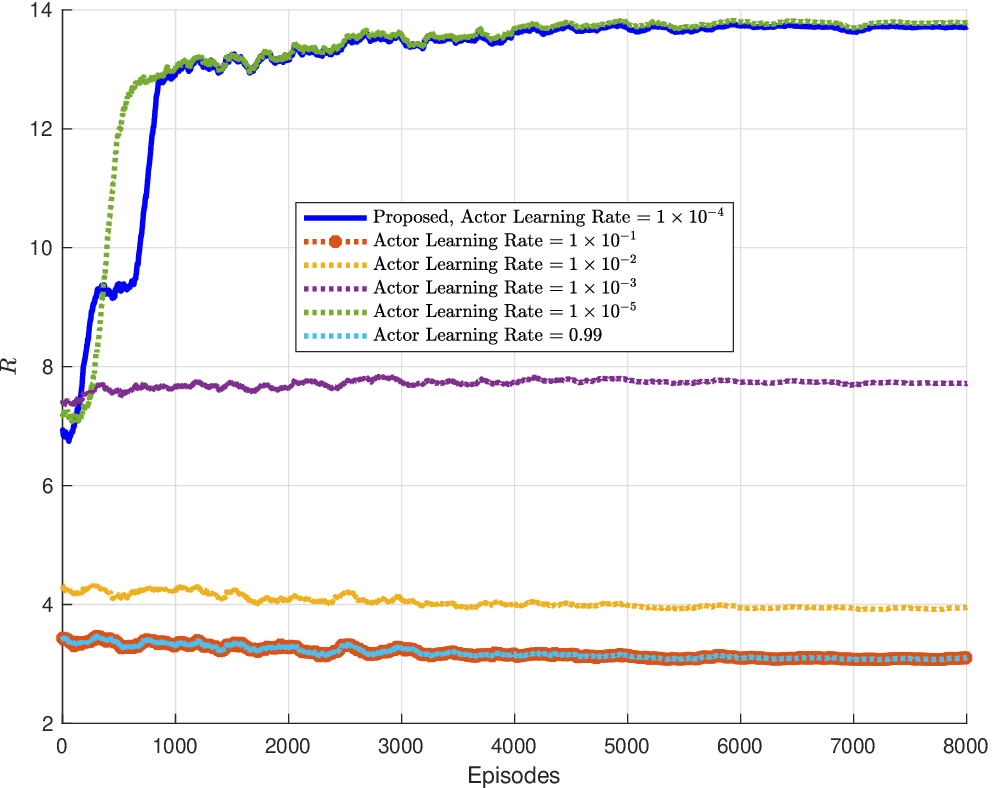}
    \caption{Learning curves of the proposed system model and solution with different actor learning rates.}
    \label{Fig1}
       \vspace{5pt}
\end{figure}

\begin{figure}[!b]
   \vspace{5pt}
    \centering
    \includegraphics[width=0.4\textwidth]{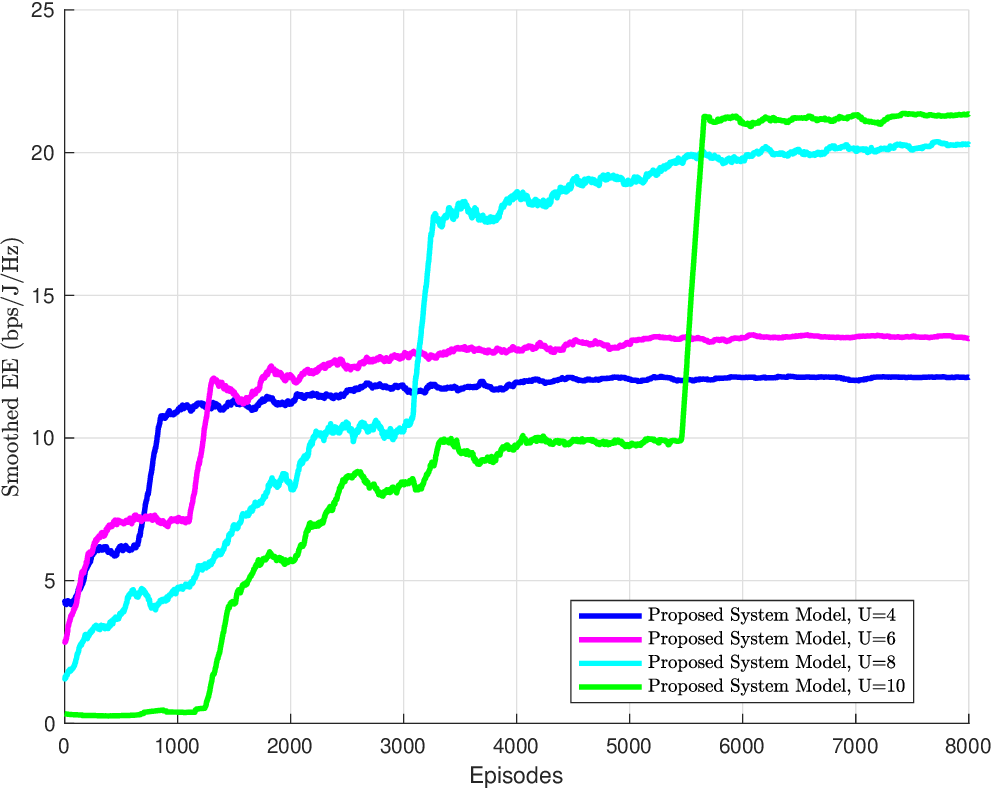}
    \caption{EE vs episodes for different number of users.}
    \label{Fig2}
       \vspace{5pt}
\end{figure}

	
	The impact of system model parameters is illustrated in Fig. \ref{Fig2}, where we evaluate the effect of increasing the number of users on the system’s EE. As shown in Fig. \ref{Fig2}, total sum rate  with the increase in the number of users, primarily due to the corresponding growth in the total sum rate. However, the convergence behavior differs across system model configurations. Specifically, as the number of users increases, convergence requires more episodes, which can be attributed to the higher complexity of the system model.

\begin{table}[t!]
    \centering
    \vspace{20pt}
    \begin{tabular}{ p{2.35cm}|p{0.9 cm}|p{2.75cm}|p{0.5cm} }
        \multicolumn{4}{c}{\textbf{Table I}: Simulation Parameters and AIW-Meta-SAC Hyper-parameters \cite{farhadi2025joint,farhadi2025meta}} \\
        \hline
        \textbf{Parameter} & \textbf{Value} & \textbf{Parameter} & \textbf{Value} \\
        \hline
        \hline
        $ U_t+U_r$ & $4$ & $M = M_x M_z$ & $4$ \\
        \hline
        $K_{\text{RIS}}$ & $4$ & $N$ & $4$ \\
        \hline
        $P_{max}$ & $10 \text{ Watts}$ & $(\sigma_u^n)^2$ & $-174 \text{ dBm/Hz}$ \\
        \hline
        $U_{\max}$ & $2$ & $L$ & $2$ \\
       \hline
        $P$ & $2$ & $\eta$ & $0.4$ \\
       \hline
        $\kappa$ & $0.03$ & $ d_x, d_z $ & $0.5$  \\
        \hline
        $P_{\text{St}}^{\text{BS}}$ & $30~\text{dBm}$ & $P_{i}^{\text{Ant}}$ & $20~\text{dBm}$ \\
 \hline
$P_{\text{St}}^{\text{BD-RIS}}$ & $100~\text{mW}$ & $P_{\text{Dn}}^{\text{BD-RIS}}$ & $0.33~\text{mW}$ \\

        \hline
        $Action Amplifier$ & $10$ & \text{mini-batch size} & $32$  \\
        \hline
        \text{Number of episodes} & $8000$ & \text{Actor learning rate} & $0.0001$ \\
        \hline
        \text{Critic learning rate} & $0.0001$ & \text{Reward discount} & $0.99$ \\
        \hline
        \text{SAC soft replacement} & $0.01$ & \text{Replay memory capacity} & $1000000$ \\
    \end{tabular}
    \label{tab:combined}
\end{table}

\begin{figure}[!t]
   \vspace{5pt}
    \centering
    \includegraphics[width=0.4\textwidth]{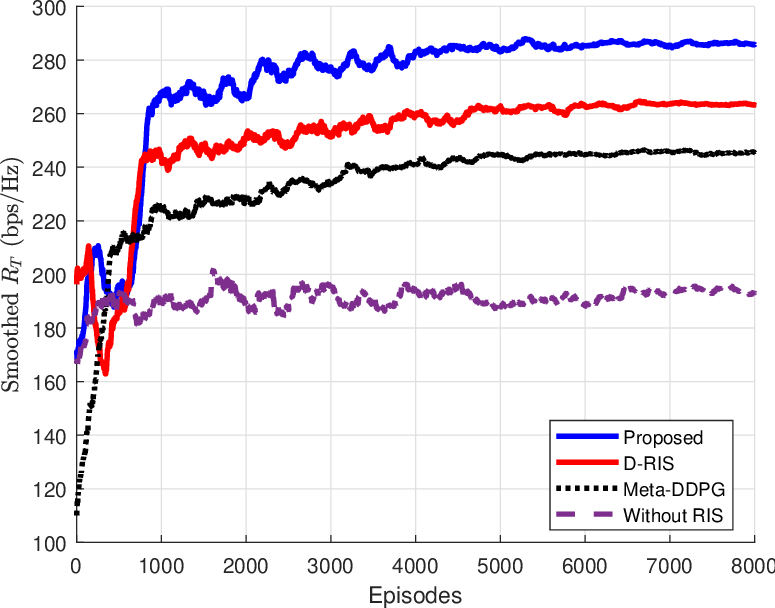}
    \caption{$R_T$ versus episode for different system models and algorithm.}
    \label{Fig3}
       \vspace{5pt}
\end{figure}

\begin{figure}[!t]
   \vspace{5pt}
    \centering
    \includegraphics[width=0.4\textwidth]{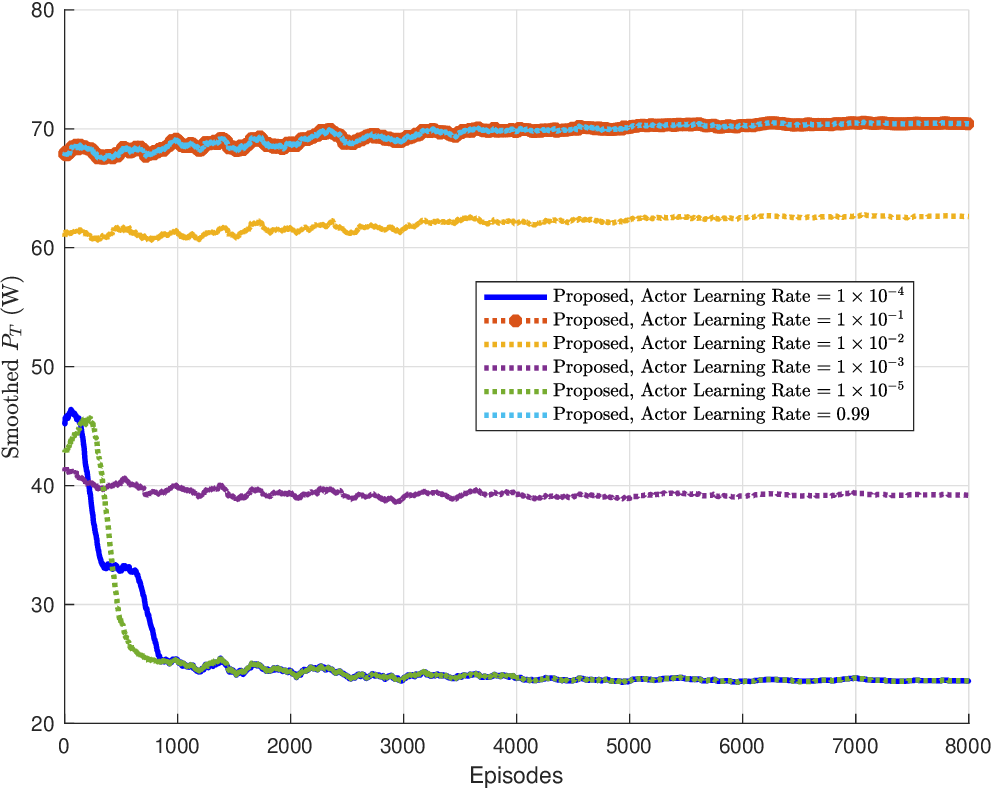}
    \caption{$P_T$ versus episodes with different actor learning rates.}
    \label{Fig4}
       \vspace{5pt}
\end{figure}

	A comparison between the proposed algorithm and Meta-DDPG, as well as the proposed system model, is presented in Fig. \ref{Fig3}. The results demonstrate that the proposed solution and system model achieve superior performance compared to the alternatives. As expected, the STAR-BD-RIS outperforms the conventional D-RIS. Furthermore, both of the aforementioned system models outperform the baseline configuration that relies solely on the FIM-equipped BS.

	 Fig. \ref{Fig4} exhibits a trend similar to that of Fig. \ref{Fig1} in terms of total power consumption. The results highlight the effect of selecting an appropriate actor learning rate on solving the optimization problem. Owing to the objective function in \eqref{OptProb}, the agent learns to enhance the EE by simultaneously reducing the total power consumption and increasing the total sum rate.
\section{Conclusion}\label{Conclusion}
This paper investigated a STAR-BD-RIS-assisted wireless communication system with a FIM-equipped BS under a NOMA framework to enhance spectral and energy efficiency. To address the resulting highly coupled and non-convex optimization problem, an AIW-Meta-SAC algorithm was proposed. The proposed approach jointly optimizes the transmit beamforming, STAR-BD-RIS configurations, NOMA parameters, and FIM shape while satisfying system constraints. Unlike conventional approaches with fixed weighting coefficients, the proposed method incorporates an AIW mechanism, which dynamically adjusts the contribution of different objective and constraint-related terms based on their historical behavior. This significantly improves learning efficiency and eliminates the need for manual tuning of weighting factors. Simulation results demonstrated that the proposed AIW-Meta-SAC framework achieves superior performance compared to benchmark methods, including Meta-DDPG and conventional system configurations such as diagonal RIS (D-RIS) and FIM-only systems. Furthermore, the impact of key system and learning parameters was analyzed. The results highlighted the critical role of the actor learning rate in ensuring stable and efficient convergence, where improper selections lead to degraded learning performance or slow convergence. Additionally, increasing the number of users improves the achievable system performance but introduces higher complexity, resulting in longer convergence times. Overall, the proposed AIW-Meta-SAC framework provides an effective and scalable solution for next-generation intelligent wireless networks, enabling efficient resource allocation and robust performance in complex environments.

\bibliographystyle{IEEEtran}
\bibliography{Bibliography}

\begin{IEEEbiography}[{\includegraphics[width=1.1in,height=1.5in, clip,keepaspectratio]{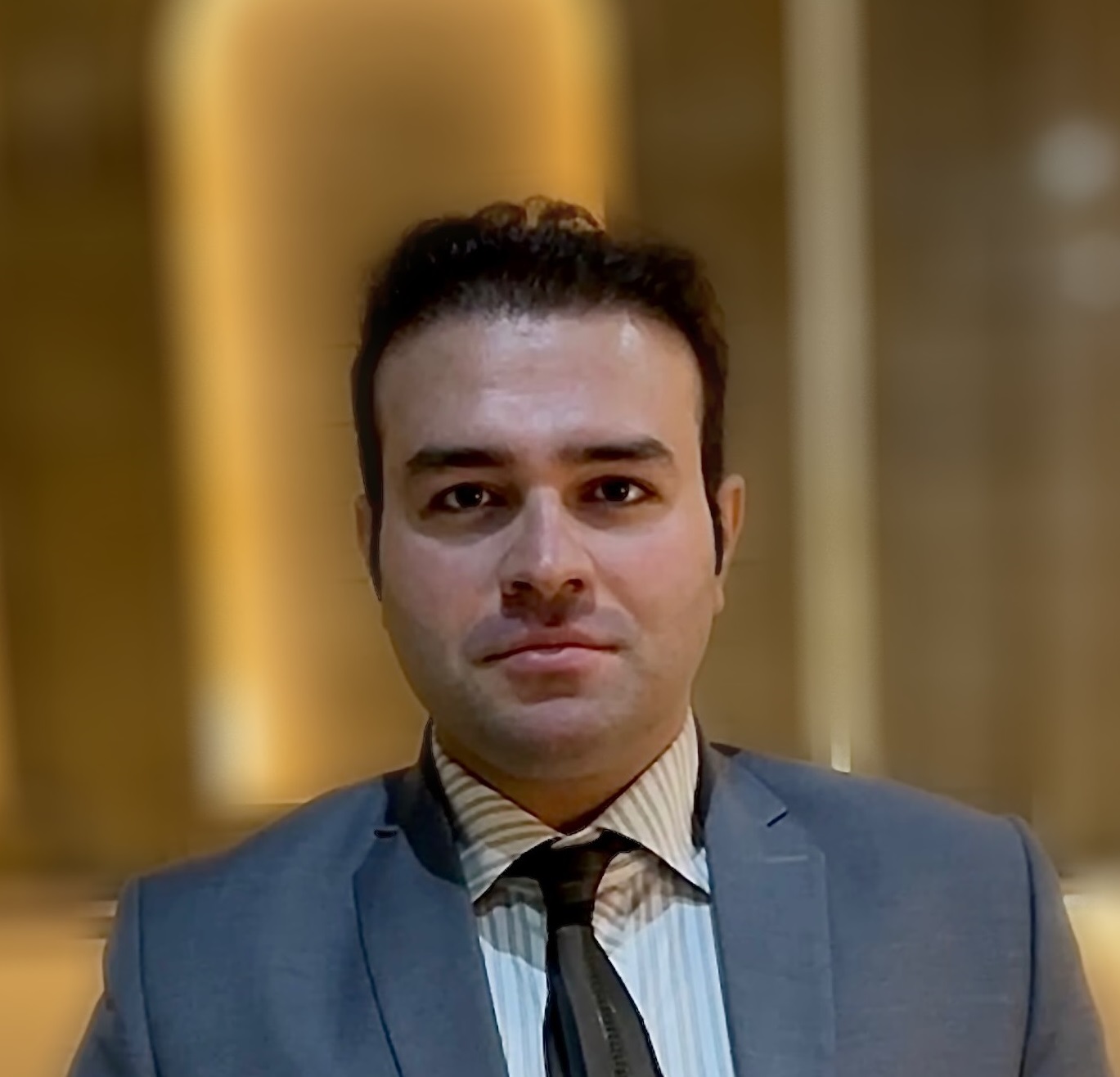}}]{Armin Farhadi} is currently a visiting Ph.D. student at UCL and a Ph.D. candidate in Electrical Engineering – Telecommunications Systems at the University of Tehran. He received his B.Sc. and M.Sc. degrees in Electrical and Telecommunications Engineering, from the University of Qom, Iran, in 2017, and in Electrical Engineering – Telecommunications Systems from Shahed University, Tehran, Iran, in 2019, respectively. He also serves as a TPC member of IEEE ICC’25 - SAC-13 ISAC Track, EEE ICC’24 - SAC-13 ISAC Track, IEEE ICC'25 - SAC-13 ISAC Track (2025 IEEE International Conference on Communications (ICC): SAC Integrated Sensing and Communication Track), 2025 5th International Conference on Computer Application and Information Security (ICCAIS 2025), 2025 IEEE 101st Vehicular Technology Conference: VTC2025-Spring, and the 2024 IEEE 35th Annual International Symposium on Personal, Indoor, and Mobile Radio Communications (PIMRC). His research interests include Machine Learning, Neural Networks, Reinforcement Learning, ISAC, IRS, STAR-IRS, Convex Optimization, Resource Allocation, 6G Networks, C-RAN, NOMA, SCMA, and Signal Processing.
\end{IEEEbiography}

\begin{IEEEbiography}[{\includegraphics[width=1.1in,height=1.5in, clip,keepaspectratio]{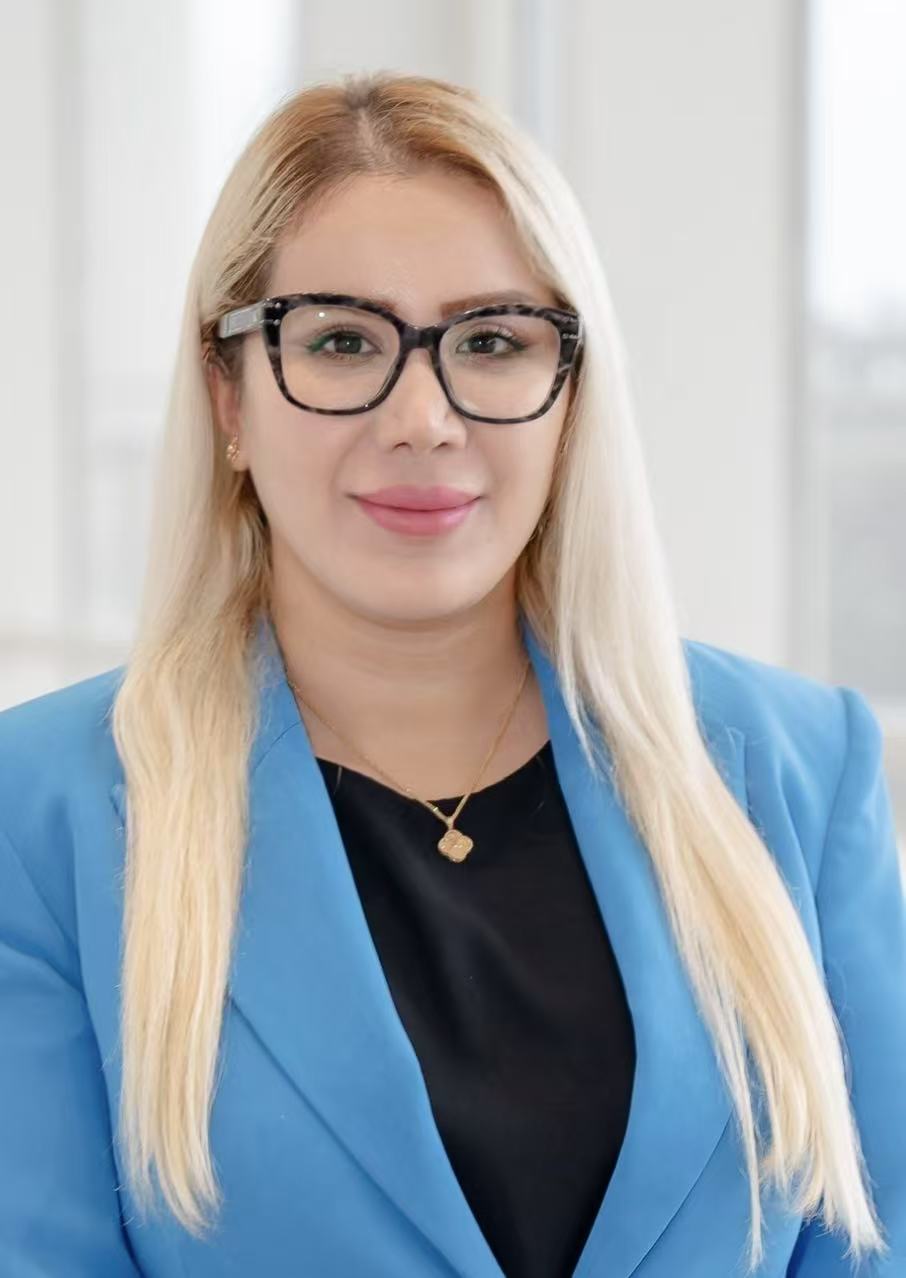}}]{Maryam Cheraghy}  received the Ph.D. degree in communication and information engineering from Shanghai Jiao Tong University, Shanghai, China. She joined the Department of Computer Science, Wenzhou-Kean University, in 2022, where she is currently a Lecturer. Her research interests include B5G/6G wireless communication, RIS, and machine learning. She is an Editorial Board Member and a reviewer of IEEE TRANSACTIONS ON VEHICU LAR TECHNOLOGY and IEEE COMMUNICATIONS LETTERS.
\end{IEEEbiography}

\begin{IEEEbiography}[{\includegraphics[width=1.1in,height=1.5in, clip,keepaspectratio]{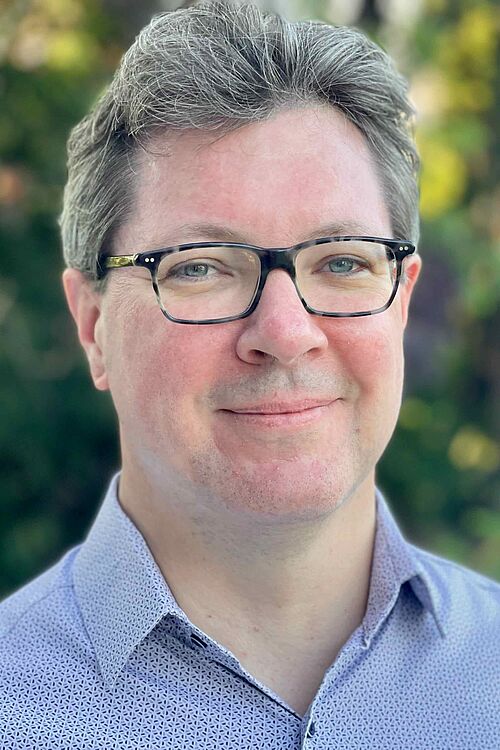}}]{Eduard A. Jorswieck} was born in 1975 in Berlin, Germany. He received his Diplom-Ingenieur (M.S.) degree and Doktor-Ingenieur (Ph.D.) degree, both in electrical engineering and computer science from the Technische Universität Berlin, Germany, in 2000 and 2004, respectively. He was with the Fraunhofer Institute for Telecommunications, Heinrich-Hertz-Institut (HHI) Berlin, in the Broadband Mobile Communication Networks Department from December 2000 to January 2008. Since April 2005 he has been a lecturer at the Technische Universität Berlin. In February 2006, he joined the Department of Signals, Sensors and Systems at the Royal Institute of Technology (KTH) as a post-doc and became a Assistant Professor in 2007. Since February 2008, he has been the head of the Chair of Communications Theory and Full Professor at Dresden University of Technology (TUD), Germany. Eduard's main research interests are in the area of signal processing for communications and networks, applied information theory, and communications theory. He has published more than 75 journal papers and some 200 conference papers on these topics. Dr. Jorswieck is senior member of IEEE. He was member of the IEEE SPCOM Technical Committee (2008-2013) and is member of the IEEE SAM Technical Committee since 2015. Since 2011, he acts as Associate Editor for IEEE Transactions on Signal Processing. Since 2008, continuing until 2011, he has served as an Associate Editor for IEEE Signal Processing Letters continuing until 2013 as Senior Associate Editor. Since 2013, he serves as Associate Editor for IEEE Transactions on Wireless Communications. In 2006, he received the IEEE Signal Processing Society Best Paper Award.
\end{IEEEbiography}

\end{document}